\documentclass[preprint,prd,aps,tighten,nofootinbib,amssymb]{revtex4}
\usepackage{color}
\input{colordvi.tex}
\usepackage{amsmath}
\usepackage[dvips]{graphicx}
\usepackage{subfigure}
\newcommand{\beq}{\begin{equation}}
\newcommand{\eeq}{\end{equation}}
\newcommand{\beqa}{\begin{eqnarray}}
\newcommand{\eeqa}{\end{eqnarray}}

\newcommand{\CO}{{\cal O}}
\newcommand{\vecs}[1]{\mbox{\boldmath${#1}$}}

\def\bu{\bar{u}}
\def\bd{\bar{d}}
\begin{document}
\widetext
\draft

\begin{flushright}
DESY 12-004
\end{flushright}

\title{Asymmetric Dark Matter from Spontaneous Cogenesis in the
Supersymmetric Standard Model}

\author{Kohei Kamada}%
\email[Email: ]{kohei.kamada"at"desy.de}
\affiliation{Deutsches Elektronen-Synchrotron DESY, Notkestra\ss e 85, D-22607 Hamburg, Germany}

\author{Masahide Yamaguchi}
\email[Email: ]{gucci"at"phys.titech.ac.jp}
\affiliation{Department of Physics, Tokyo Institute of Technology, Tokyo 152-8551, Japan}

\date{\today}


\begin{abstract}
The observational relation between the density of baryon and dark matter
in the Universe, $\Omega_{\rm DM}/\Omega_B\simeq 5$, is one of the most
difficult problems to solve in modern cosmology. We discuss a scenario
that explains this relation by combining the asymmetric dark matter
scenario and the spontaneous baryogenesis associated with the flat
direction in the supersymmetric standard model. A part of baryon
asymmetry is transferred to charge asymmetry $D$ that dark matter
carries, if a symmetry violating interaction that works at high
temperature breaks not only $B-L$ but also $D$ symmetries
simultaneously. In this case, the present number density of baryon and
dark matter can be same order if the symmetric part of dark matter
annihilates sufficiently. Moreover, the baryon number density can be
enhanced as compared to that of dark matter if another $B-L$ violating
interaction is still in thermal equilibrium after the spontaneous
genesis of dark matter, which accommodates a TeV scale asymmetric dark
matter model.
\end{abstract}

\maketitle

\section{Introduction}

The existence of dark matter (DM) \cite{arXiv:1001.4538,hep-ph/0404175}
and the present baryon asymmetry \cite{Sakharov:1967dj, arXiv:1001.4538}
are the most important challenges in modern cosmology since it is quite
difficult to accommodate them in the context of the standard model of
particle physics (SM). Many attempts to explain them separately have
been proposed. For example, thermal relics of weakly interacting massive
particles (WIMP) can explain the present dark matter abundance elegantly
\cite{hep-ph/0404175, kolbturner}. Affleck-Dine (AD) baryogenesis
\cite{Affleck:1984fy, Dine:1995kz}, baryogenesis through leptogenesis
\cite{RIFP-641}, electroweak baryogenesis \cite{hep-ph/9302210}, and so
on have been proposed as viable models for baryogenesis. However, the
coincidence of the present density parameter of baryons and dark matter,
$\Omega_{\rm DM}/\Omega_B\simeq 5$ \cite{arXiv:1001.4538}, is extremely
difficult to explain. One of the reasons lies in the fact that $CP$
violation is essential for baryogenesis \cite{Sakharov:1967dj} while
the thermal relic DM does not need $CP$ violation. Thus, we cannot help but
regard such a coincidence as an accident as far as the thermal relic
scenario is responsible for the present dark matter
abundance.\footnote{Very recently, an interesting WIMP scenario
\cite{Cui:2011ab} was proposed, in which WIMP dark matter annihilation
is directly responsible for baryogenesis.}

Recently, an alternative scenario of DM that explains this coincidence
dubbed as ``asymmetric dark matter (ADM)'' \cite{arXiv:0901.4117} has
been paid great attention.  In this scenario, DM is assumed to be
charged under a global symmetry, which guarantees the stability of
DM. Asymmetry between DM and ``anti''-DM is generated by the same
physical origin as baryon asymmetry or transferred from baryon asymmetry
generated by baryogenesis mechanism \cite{ADM}. If DM annihilation works
sufficiently, the present DM is fully asymmetric so that the DM
abundance coincides with its asymmetry. Since the present number
densities of baryons and DM are linked (typically almost the same) in
this scenario, the coincidence is naturally explained as long as the
mass of DM, $M_{\rm DM}$, is of the order of GeV scale.\footnote{See,
for example, Ref.~\cite{other} for other attempts to explain the
coincidence.} In fact, the mass of DM is related to the proton mass,
$m_{\rm p}$, as follows: 
\begin{equation}
M_{\rm DM}= \frac{\Omega_{\rm DM}}{\Omega_B}\frac{n_B/s}{n_{\rm
 DM}/s}m_{\rm p} = {\cal O}(1-10) m_{\rm p}, 
\end{equation}
where $n_{B({\rm DM})}/s$ and $\Omega_{B({\rm DM})}$ represent the
baryon (DM) number-to-entropy ratio and the energy density parameter of
baryon (DM), respectively.

Among many baryogenesis models, spontaneous baryogenesis
\cite{Cohen:1987vi} is an attractive scenario in that it can work even
in thermal equilibrium.  In this mechanism, a nonvanishing velocity of
a scalar field, ${\dot \phi}$, violates $CPT$ invariance of the system and
generates an effective chemical potential between baryon and antibaryon
through a derivative coupling between the scalar field and the baryon
current. When the system is in thermal equilibrium including baryon
symmetry breaking interactions, the distribution of baryon and
antibaryon differs due to the chemical potential, which implies the
generation of baryon asymmetry even without $CP$ violation. Such baryon
asymmetry is fixed when the symmetry breaking interaction freezes
out. Thus, as long as a slow-rolling scalar field derivatively coupled
to the baryon/lepton current, it is easy to realize baryogenesis
\cite{spontaneous}.

An ADM model associated with spontaneous baryogenesis dubbed as
``spontaneous cogenesis'' has been proposed by March-Russell and
McCullough recently \cite{arXiv:1106.4319}. They discussed the ADM
scenario by introducing a global $U(1)_X$ symmetry and a slow-rolling
light scalar field, which is assumed to be derivatively coupled to the
$X$ current. The asymmetry is generated in the DM sector and transmitted
to the visible sector through the mixing operator which mediates $X$ and
$B-L$ violating interactions but preserves one of their
combinations. The annihilation of the symmetric part of dark matter
works well in some region of the parameter space. However, since the
spontaneous mechanism works in the DM sector, the origins of not only
the DM field but also the light scalar field are unidentified, in
particular, the presence of the derivative coupling is simply assumed.

On the other hand, Chiba, Takahashi, and one of the present authors (MY)
have pointed out that such a derivative interaction can be naturally
realized and spontaneous mechanism works well
\cite{arXiv:hep-ph/0304102} in the context of flat directions in the
minimal supersymmetric standard model (MSSM) \cite{hep-ph/9510370}. Once
a flat direction acquires the vacuum expectation value (VEV), the
symmetry possessed by the flat direction, typically a combination of the
$B$ and $L$ symmetries, is broken. Then, the Nambu-Goldstone (NG) boson
associated with its symmetry breaking is shown to be derivatively
coupled to the current. Its slow-roll motion due to an explicit symmetry
breaking $A$ term induces the chemical potential of particles charged
under the symmetry. Thus, baryon ($B-L$) asymmetry is spontaneously
generated as long as the symmetry breaking/mixing operator, which can be
easily introduced as a nonrenormalizable operator in the MSSM sector,
is in thermal equilibrium. Moreover, it is shown that this mechanism can
work even for a flat direction without $B-L$ charge \footnote{For the
flat direction with $B-L$ charge, AD mechanism works
effectively as well as spontaneous baryogenesis. It should be also
noticed that this spontaneous mechanism works for a flat direction with
neither baryon nor lepton charge by introducing another charge like the
Peccei-Quinn charge.\cite{arXiv:hep-ph/0304102}} thanks to the mixing
operator, which is essentially the same introduced later by
March-Russell and McCullough in the context of cogenesis.

In this paper, we propose another ADM scenario by use of the MSSM flat
direction, in which the associated NG boson derivatively couples to a
combination of the $B$ and $L$ currents, and its velocity induces an
effective chemical potential. By introducing a DM field, its associated
charge $D$, and a nonrenormalizable mixing interaction that violates
$B, L$ and $D$ charges simultaneously, dark matter and $B-L$ asymmetries
are generated at the same time, which favors dark matter with GeV scale
mass. Moreover, typically speaking, there also exists an interaction
violating only a combination of $B$ and $L$ charges in the MSSM sector.
Provided that such an interaction freezes out a little after the
freeze-out of the mixing interaction,\footnote{This requirement is
natural because the latter interaction breaks the symmetry more strongly
than the former interaction.} the baryon number density becomes slightly
larger than that of the DM. Thus, our scenario can accommodate DM with
the weak-scale mass, which may enable us to easily identify the origin
of DM.\footnote{In the case that a light scalar field derivatively
couples to the $D$ current as discussed in Ref. \cite{arXiv:1106.4319},
such a slight enhancement happens for the number density of dark matter
instead of baryon charge, even if we introduce an interaction violating
only $D$ charges. In this case, the mass of dark matter must be smaller
than GeV.} We also find that the annihilation of the symmetric part of
DM works efficiently if we extend our visible sector to the
next-to-minimal supersymmetric standard model (NMSSM)
\cite{arXiv:0906.0777}.  Thus, our scenario is suitable for realizing an
ADM model and even accommodates TeV scale dark matter.

This paper is organized as follows. In the next section, the spontaneous
baryogenesis in the MSSM flat direction is briefly reviewed. A general
discussion of the spontaneous cogenesis as an ADM model is also made in
the context of the MSSM. In Sec.~\ref{sec:3}, we discuss concrete
examples of spontaneous cogenesis and investigate realistic parameter
range to explain the present baryon and DM abundance simultaneously. We
also comment on the annihilation process of symmetric part of
DM. Section \ref{sec:4} is devoted to our conclusions and a discussion.

\section{Spontaneous Cogenesis in a flat direction}

\subsection{Review of Spontaneous Baryogenesis in a Flat Direction \label{sec:2.1}}

First we briefly review the spontaneous baryogenesis mechanism proposed
by Cohen and Kaplan \cite{Cohen:1987vi} and its realization in a flat
direction of the supersymmetric standard model
\cite{arXiv:hep-ph/0304102}. Let us consider an effective Lagrangian in
which a scalar field $a$ derivatively couples to the baryon current
$J^\mu_B$,\footnote{Here we consider only the baryon symmetry $U(1)_B$
for simplicity but it can be straightforwardly extended to the case with
another global $U(1)$ symmetry.}
\begin{equation}
{\cal L}_{\rm eff} = -\frac{\partial_\mu a}{M} J^\mu_B  \label{Leff}.
\end{equation}
Here $M$ is a cutoff scale and the baryon current is given by
\begin{align}
J^\mu_B &=\sum_m B_m j^\mu_{m}, \notag \\
j^\mu_{m}&={ \left\{ 
\begin{array}{ll}
{\bar \psi}_m \gamma^\mu \psi_m & \quad \text{for fermions}, \\
i(\varphi_m \partial_\mu \varphi_m^*-\varphi_m^* \partial^\mu \varphi_m) & \quad \text{for complex scalar fields}, 
\end{array} 
\right. }
\end{align}
with $B_m$ being the baryon number of the field $\psi_m (\varphi_m)$.
When a homogeneous scalar field $a$ acquires a nonvanishing classical
velocity, $\partial_\mu a =({\dot a}, {\vecs 0})$, the effective
Lagrangian \eqref{Leff} reads,
\begin{equation}
{\cal L}_{\rm eff} = - \sum_m \frac{\dot a}{M} B_m n_m = - \frac{\dot a}{M} n_{B}, 
\end{equation}
where we have used the fact that the $0$-th component of baryon current
represents the baryon number density, $\sum_m B_m j^0_m=\sum_m B_m
n_m=n_{B}$. This effective Lagrangian can be regarded as number density
multiplied by the effective chemical potential $\mu_m \equiv -{\dot a}
B_m/M$. If there is a baryon number violating interaction and the
Universe is hot enough for the interaction to be in thermal equilibrium,
fields $\psi_m (\varphi_m)$ with baryonic charges are distributed
according to their chemical potentials, which leads to the generation of
baryon asymmetry in the Universe. Note that this process does not
require the Sakharov's condition. Instead, a nonvanishing classical
value of ${\dot a}$ breaks the $CPT$ invariance since ${\dot a}$ is odd
under the $CPT$ transformation, which spontaneously generates baryon
asymmetry.

As the temperature of the Universe decreases, the baryon number
violating interaction decouples at $T=T_{\rm dec}$. If it happens before
the decay of the field $a$, the baryon number density is frozen out at
the value,
\begin{equation}
n_B(t_{\rm dec}) = \sum_m B_m\frac{g_m \kappa_m T_{\rm dec}^3}{6} \left(\frac{\mu_m}{T_{\rm dec}}+\CO\left[ \left(\frac{\mu_m}{T_{\rm dec}}\right)^3\right]\right), 
\end{equation}
where $g_m$ is the degree of freedom of $\psi_m(\varphi_m)$ and
$\kappa_m$ is defined as
\begin{equation}
\kappa_m =\left\{
\begin{array}{ll}
1 & ~~\text{for fermions }(\psi_m), \\
2 & ~~\text{for scalar fields }(\varphi_m). 
\end{array}\right. 
\end{equation}
The baryon-to-entropy ratio is fixed at reheating if $T_{\rm dec}>T_R$
and at the decoupling temperature if $T_R>T_{\rm dec}$, where $T_R$ is
the reheating temperature. Before reheating, the Universe is
dominated by the inflaton oscillation, but thermal plasma has already
existed, during which the temperature is related to the Hubble parameter
as $T\simeq (H M_G T_R^2)^{1/4}$ \cite{kolbturner} with $M_G$ and $H$
being the reduced Planck mass and the Hubble parameter,
respectively. Thus, the present baryon-to-entropy ratio is given by
\begin{equation}
\frac{n_B}{s}\simeq \left\{
\begin{array}{ll}
\dfrac{15}{4 \pi^2 g_{*s}} \sum_m B_m \dfrac{g_m \kappa_m \mu_m}{T_{\rm
 dec}}& ~~~\text{for }~ T_{\rm dec}<T_R, \\
\dfrac{15}{4 \pi^2 g_{*s}} \sum_m B_m \dfrac{g_m \kappa_m \mu_m}{T_{\rm
 dec}} \left(\dfrac{T_R}{T_{\rm dec}}\right)^5& ~~~\text{for }~ T_{\rm dec}>T_R,
\end{array}
\right.
\end{equation}
where $g_{*s}\simeq 200$ is the effective degrees of freedom of
relativistic fields. No entropy production after reheating is also
assumed.

Following the discussion of Ref.~\cite{arXiv:hep-ph/0304102}, we see how
the effective Lagrangian like Eq.~\eqref{Leff} naturally arises and
spontaneous baryogenesis is realized in the context of the
supersymmetric standard model. In supersymmetric theories, there are
many flat directions along which the scalar potential vanishes. Since a
flat direction is charged under the $U(1)_B$ and/or $U(1)_L$ symmetries,
such symmetries are spontaneously broken if scalar fields acquire
nonvanishing expectation values along the flat direction.  Then, the
NG boson associated with this symmetry breaking
derivatively couples to its currents, which is exactly the effective
interaction we want. More concretely, a flat direction can be
parameterized by composite holomorphic gauge-invariant polynomials as
\begin{equation}
X=\prod_{i=1}^N \chi_i, 
\end{equation}
where $N$ is the number of superfields $\chi_i$ that constitute the flat
direction. The expectation value of a scalar field corresponding to
$\chi_i$ (we use the same symbol for a superfield and its scalar part)
can be decomposed as
\begin{equation}
\langle \chi_i \rangle =\frac{f_i}{\sqrt{2}} e^{i \theta_i}, 
\label{eq:chii}
\end{equation}
where $f_i/\sqrt{2}$ is the absolute value of $\chi_i$ and $\theta_i$ is
its phase. Note that their values are related each other due to the $F$-
and $D$-flat conditions. Since the field $\chi_i$ is charged under the
$U(1)_B$ or $U(1)_L$ symmetry, we need to treat both symmetries
adequately for spontaneous baryogenesis in a flat direction. For this
purpose, we consider $U(1)_{A^\pm}$ symmetries, which are the two
independent linear combinations of the $U(1)_B$ and $U(1)_L$
symmetries. The $U(1)_{A^\pm}$ charges of the field $\chi_i$ are defined
as
\begin{align}
Q^+_i &= B_i \cos \xi + L_i \sin \xi \quad~~~\, \text{for }~~ U(1)_{A^+},  \label{qplus}\\
Q^-_i &=-B_i \sin \xi +L_i \cos\xi \quad~ \text{for }~~ U(1)_{A^-}, \label{qminus}
\end{align}
with $\xi$ given by
\begin{equation}
\tan \xi \equiv \frac{\sum_{i=1}^N L_i}{\sum_{i=1}^N B_i}. \label{xi}
\end{equation}
Here we denote the charges of $\chi_i$ as $B_i$ and $L_i$, respectively.
The flat direction $X$ is charged under $U(1)_{A^+}$ but is not charged
under $U(1)_{A^-}$ as a whole.

In order to identify NG bosons, we express the phase of the scalar field
$\theta_i$ as \cite{sikivie,Dolgov:1994zq}
\begin{equation}
\theta_i=B_i \alpha_B+L_i \alpha_L = Q^+_i \alpha_+ + Q^-_i \alpha_-, 
\end{equation}
where $\alpha_B, \alpha_L, \alpha_+$ and $\alpha-$ are the angles
conjugate to the generators of the symmetries, respectively, and are
related as 
\begin{equation}
\begin{pmatrix}
\alpha_B \\ \alpha_L
\end{pmatrix}=\begin{pmatrix}
\cos \xi & -\sin \xi \\
\sin \xi & \cos \xi 
\end{pmatrix} 
\begin{pmatrix}
\alpha_+ \\ \alpha_-
\end{pmatrix}. 
\end{equation}
The NG bosons associated with the spontaneous breaking of the
$U(1)_{A^\pm}$ symmetries are denoted as $a_{\pm}$ and can be expressed
by the angles $\alpha_{\pm}$ through the decay constant matrix $F$,
\begin{equation}
\begin{pmatrix}
a_+ \\ a_-
\end{pmatrix}
=F 
\begin{pmatrix}
\alpha_+ \\ \alpha_-
\end{pmatrix}
.
\end{equation} 
Since the (canonical) kinetic terms of $a_{\pm}$ are given by those of
$\chi_{i}$ with $f_i$ fixed, the decay constant matrix $F$ satisfies the
following relation: 
\begin{equation}
F^T F = \sum_i f_i^2
\left(
\begin{array}{cc}
 Q_i^+ Q_i^+&Q_i^+ Q_i^-\\
 Q_i^- Q_i^+&Q_i^- Q_i^-\
\end{array}
\right).
\label{eq:Fcond}
\end{equation}
On the other hand, the decay constant matrix $F$ takes the following
form because only the NG boson $a_{+}$ is lifted by an $A$ term, as we
will see later: 
\begin{equation}
F =
\begin{pmatrix}
v_a & 0 \\ f_{01} & f_{11}
\end{pmatrix}.
\label{eq:F2}
\end{equation}
Combining Eqs. (\ref{eq:Fcond}) and (\ref{eq:F2}) yields the decay
constants,
\begin{equation}
v_a^2=\sum_if_i^2 Q_i^{+2}-\frac{\left(\sum_i f_i^2 Q_i^+ Q_i^- \right)^2}{\sum_i f_i^2 Q_i^{-2}}, \quad f_{01}=\frac{\sum_i f_i^2 Q_i^+ Q_i^-}{\sqrt{\sum_i f_i^2 Q_i^{-2}}}, \quad f_{11}=\sqrt{\sum_i f_i^2 Q_i^{-2}}. \label{decayconst}
\end{equation}
A flat direction is often parameterized by a representative scalar field
$\Phi$ as
\begin{equation}
X=\Phi^N,
\end{equation}
where 
\begin{equation}
\Phi=\frac{\phi}{\sqrt{2}}e^{i \theta}, \quad \phi\equiv \left(\prod_i f_i\right)^{1/N}, \quad \theta\equiv\frac{1}{N}\sum_i \theta_i=\frac{a_+}{N v_a} Q^+,  
\end{equation}
with $Q^+ \equiv \sum_i Q^+_i=\pm \sqrt{(\sum_i B_i)^2+(\sum_i L_i)^2}$
depending on the sign of $\cos{\xi}$. Notice that $a_-$ does not appear
here because the flat direction is not charged under the $U(1)_{A^-}$
symmetry as a whole.

Now we derive a derivative coupling of $a_+$ and the $U(1)_{A^+}$
current as a consequence of the spontaneous breaking of the $U(1)_{A^+}$
symmetry due to the nonvanishing expectation values of scalar fields
along a flat direction.\footnote{There is also a derivative coupling
between $a_-$ and the $U(1)_{A^-}$ current, but $\partial_\mu a_-$ does
not acquire a nonvanishing classical value because it remains massless
even with the presence of an $A$ term.} Under the infinitesimal
$U(1)_{A^+}$ transformation, the NG boson $a_+$ transforms as $a_+
\rightarrow a_+ + v_a \epsilon$, where $\epsilon$ is the infinitesimal
transformation parameter.  At the same time, a matter field $\chi_m$
that has a $U(1)_{A^+}$ charge $Q_m^+$ transforms as $\chi_m \rightarrow
\chi_m+ \epsilon \delta \chi_m$ with $\delta \chi_m=i Q_m^+
\chi_m$. Then, the $U(1)_{A^+}$ current defined as
\begin{equation}
J_{A^+}^\mu\equiv -\sum_{m^\prime} \frac{\partial {\cal L}}{\partial (\partial_\mu \chi_{m^\prime})}\delta \chi_{m^\prime}, 
\end{equation}
where $m^\prime$ runs all the fields that are charged under the
$U(1)_{A^+}$ symmetry, reduces to
\begin{equation}
J^\mu_{A^+} = v_a \partial^\mu a_+ + \sum_m Q_m^+  j_m^\mu. 
\end{equation}
Current conservation gives the equation of motion for $a_+$,
\begin{equation}
\partial_\mu J^\mu_{A^+} = v_a \partial^2 a_+ + \sum_m Q_m^+ \partial_\mu  j_m^\mu =0. \label{Curcon}
\end{equation}
While the first term in the middle equation in Eq. \eqref{Curcon}
represents the kinetic term for the NG mode $a_+$, the second term comes
from the following effective Lagrangian: 
\begin{equation}
{\cal L}_{\rm eff}=-\sum_m \frac{Q_m^+}{v_a} (\partial_\mu a_+) j_m^\mu, \label{Leffm}
\end{equation}
which represents a derivative coupling of the NG boson $a_{+}$ and the
charged fields. Thus, a nonvanishing velocity of $a_{+}$ can lead to
spontaneous genesis of $U(1)_{A^+}$ asymmetry as long as the symmetry
breaking operator is in thermal equilibrium. When the field $a_+$ is
homogeneous, the effective Lagrangian reduces to ${\cal L}_{\rm
eff}=\sum_m \bar{\mu}_m n_m$ with the chemical potential
$\bar{\mu}_m$ given by
\begin{equation}
  \bar{\mu}_m \equiv - \frac{Q_m^+}{v_a} \dot{a}_{+}.
  \label{eq:effchem}
\end{equation}

Next, let us discuss the dynamics of a flat direction. Though the scalar
potential vanishes along the flat direction in the supersymmetric and
renormalizable limit, it can be lifted \cite{Dine:1995kz} by the supersymmetry (SUSY)-breaking effects and
a nonrenormalizable superpotential of the form, 
\begin{equation}
W_{\rm NR} =\frac{X^k}{Nk M_*^{Nk-3}} =\frac{\Phi^n}{nM_*^{n-3}},
\end{equation}
with $M_*$ being a cutoff scale and $n=Nk$. Then, the resultant scalar
potential is written as \cite{Dine:1995kz}
\begin{equation}
V=V_{\not {\rm S}} + \left(a_{3/2}\frac{m_{3/2}}{nM_*^{n-3}}\Phi^n+{\rm h.c.}\right)+\frac{|\Phi|^{2n-2}}{M_*^{2n-6}}. 
\end{equation}
Here $V_{\not{\rm S}}$ represents the soft breaking effect that depends
on the SUSY-breaking mechanism, and the second term in the right-hand side
comes from supergravity effects and is called an $A$ term.  $m_{3/2}$ is
the gravitino mass and $a_{3/2}$ is a complex numerical factor whose
amplitude is of the order of unity.\footnote{Since $a_{3/2}$ can be real
by field redefinition, we treat it as a real parameter.}

During inflation, a flat direction can acquire negative Hubble induced
mass squared if it has a noncanonical interaction to the inflaton in
the K{\"a}hler potential with appropriate sign and magnitude
\cite{Dine:1995kz},
\begin{equation}
V_H=-c_H H^2 |\Phi|^2, \label{hmass}
\end{equation}
where $c_H$ is a positive constant of the order of unity. This term
destabilizes the flat direction from the origin so that it acquires a
large expectation value, whose magnitude is determined by the balance
between the negative Hubble induced mass term and the $F$ term,
\begin{equation}
\phi_{\rm min} \simeq (H M_*^{n-3})^{1/(n-2)}. 
\label{eq:phim_hubble}
\end{equation}
During the inflaton oscillation dominated era, in addition to the Hubble
induced mass \eqref{hmass} which remains until the inflaton decay (that
is, reheating), thermal effects \cite{Anisimov:2000wx} can also kick the
flat direction from the origin through the negative thermal logarithmic
potential \cite{Kasuya:2003yr},\footnote{Even during the inflaton
oscillation dominated era, thermal plasma exists as a subdominant
component of the Universe, coming from the inflaton decay.}
\begin{equation}
V_{\rm thlog}=-\alpha T^4 \log \left(\frac{|\Phi|^2}{T^2}\right). \label{negthlog}
\end{equation}
This term represents a two-loop finite temperature effect coming from
the running of the gauge/Yukawa coupling associated with the nonzero
expectation value of the flat direction. Though the sign of $\alpha$
depends on a flat direction, we take it to be positive here. Depending
on which term dominates, the flat direction sits on the value given in
\eqref{eq:phim_hubble} or the minimum given by
\begin{equation}
\phi_{\rm min} \simeq (\alpha^{\frac12} T^2 M_*^{n-3})^{1/(n-1)}. 
\end{equation}

As the Hubble parameter and the temperature of the Universe decrease,
the soft SUSY-breaking mass term and other zero temperature terms
overwhelm these terms, which makes the potential minimum the origin.
The radial component of the flat direction, $\phi$, starts its
oscillation around the origin when $H^2 \lesssim \partial^2
V/\partial\phi^2$.  In the same way, the NG boson also starts its
oscillation when
\begin{equation}
H^2 \lesssim V^{\prime\prime}, \label{SR}
\end{equation}
where $V^{\prime\prime}$ represents the second derivative of the scalar
potential with respect to $a_{+}$. The oscillation of the NG boson
$a_{+}$ typically takes place earlier than that of $\phi$. Until the
onset of its oscillation, the NG boson remains in the slow-roll regime,
which is required for the spontaneous genesis \cite{Dolgov:1994zq}.

The motion of the NG boson $a_+$ is governed by an $A$ term because it
breaks the $U(1)_{A^+}$ symmetry explicitly and generates the potential
for $a_{+}$. The $A$ term can be expressed as
\begin{equation}
V_A=\frac{a_{3/2} m_{3/2} }{2^{\frac{n}{2}-1}n M_*^{n-3}} \phi^n \cos[n
 \theta]
= M_A^4 \cos\left[\frac{k Q^+}{v_a} a_+\right], 
\end{equation}
with $M_A^4 \equiv \phi^n a_{3/2} m_{3/2} /
(2^{n/2-1}\,n\,M_*^{n-3})$. Then, the equation of motion of $a_+$ is
given by
\begin{equation}
{\ddot a}_+ + 3 H {\dot a}_+ -M_A^4 \frac{k Q^+}{ v_a} \sin\left[\frac{k Q^+}{v_a} a_+\right]=0. 
\end{equation}
While the condition $V^{\prime \prime} \simeq m_{3/2} M_A^4/\phi_{\rm
min}^2 \lesssim H^2$ is met, the NG boson $a_{+}$ is in the slow-roll
regime, which leads to
\begin{equation}
3 H {\dot a}_+ + M_A^4 \frac{k Q^+}{v_a} \simeq 0.
\end{equation}
Here we have approximated $\sin(k Q^+ a_+/v_a) \simeq 1$ and $\phi
\simeq \phi_{\rm min}$. Then, ${\dot a}_+$ acquires a nonvanishing
expectation value,
\begin{equation}
|{\dot a}_+| \simeq \frac{kQ^+}{H v_a} M_A^4. \label{adot1}
\end{equation}
For successful spontaneous baryogenesis, the $B-L$ breaking interaction
must be decoupled during the slow-roll of the NG boson $a_+$ because,
otherwise, the resulting $B-L$ asymmetry is severely suppressed
\cite{Dolgov:1994zq}.

Let us see how spontaneous baryogenesis works in a flat direction. Since
the sphaleron process is expected to work at a later epoch,\footnote{The
sphaleron process is usually assumed to be in thermal equilibrium for $T
\lesssim 10^{12}$ GeV. However, in case that the weak gauge bosons
become massive due to the large VEV of a flat direction, the sphaleron
configuration is not excited so that it is effective only after the
decay of the flat direction.} $B-L$ asymmetry must be generated for
successful baryogenesis. Otherwise, the sphaleron process would wash out
both $B$ and $L$ asymmetries. For this purpose, we consider a $B-L$
breaking interaction, whose amount of $B$ and $L$ violations are denoted
as $\Delta_B$ and $\Delta_L$, respectively.  Since $B$ and $L$
asymmetries are generated only through this interaction, they must
satisfy the following relation: 
\begin{equation}
  \frac{n_B}{\Delta_B}=\frac{n_L}{\Delta_L}.
  \label{eq:BL}
\end{equation}
However, the asymmetries based on the chemical potential
$\bar{\mu}_m$ defined in Eq.~(\ref{eq:effchem}) do not necessarily
obey the above constraint, which requires the chemical potential to be
modified. The constraint \eqref{eq:BL} is rewritten in terms of the
number density of $\chi_m$, $n_m$, as
\begin{equation}
\sum_m \Xi_m n_m=0, \quad \Xi_m\equiv B_m\Delta_L-L_m\Delta_B, \label{constfor0}
\end{equation}
which is also equivalent to the condition for the chemical potential for
$\chi_m$, $\mu_m$,
\begin{equation}
\sum_m \kappa_m g_m \Xi_m \mu_m=0. 
\end{equation}
Thus, ${\bar \mu}_m$ should be modified so as to be perpendicular to
$\Xi_m$, yielding the physical chemical potential $\mu_m$,
\begin{equation}
\mu_m = {\bar \mu}_m-\frac{({\bar \mu} \cdot \Xi)}{\Xi^2}\Xi_m, \label{chemm}
\end{equation}
where we have defined the following shorthand: 
\begin{equation}
Y^2 \equiv \sum_m \kappa_m g_m Y_m^2, \quad Y \cdot Z \equiv \sum_m \kappa_m g_m Y_m Z_m. 
\end{equation}
Since the number density of $\chi_m$, $n_m$, at the decoupling
temperature is expressed as
\begin{equation}
n_m(T_{\rm dec})=\frac{\kappa_m g_m}{6} \mu_m T_{\rm dec}^2,
\end{equation}
the resultant $B-L$ asymmetry is given by
\begin{align}
n_{B-L}(T_{\rm dec})=\sum_m (B_m-L_m) n_n(T_{\rm dec}) = (\Delta_B-\Delta_L) (\mu_B\Delta_B+\mu_L\Delta_L)\frac{B^2L^2}{B^2 \Delta_L^2+L^2\Delta_B^2}\frac{T_{\rm dec}^2}{6}. \label{bln1}
\end{align}
Here we have defined 
\begin{equation}
\mu_B \equiv -\frac{{\dot a}_+ \cos \xi}{v_a}, \quad \mu_L  \equiv -\frac{{\dot a}_+ \sin \xi}{v_a}, 
\end{equation}
and assumed $B_m L_m=0$ for all the $\chi_m$ fields, which are
reasonable for the (supersymmetric) standard model particles. From
\eqref{bln1}, it is manifest that the following two conditions must be
satisfied for successful $B-L$ genesis: 
\begin{equation}
\Delta_B - \Delta_L \not= 0, \quad 
\mu_B \Delta_B+\mu_L\Delta_L = - \Delta_{Q^{+}} \frac{\dot{a_{+}}}{v_a}
 \not=0. \label{spobarcond}
\end{equation}
Thus, the $B-L$ asymmetry is generated even if a flat direction $X$
itself does not have a $B-L$ charge. Depending on whether the $B-L$
breaking interaction is decoupled before or after the reheating, the
final $B-L$ asymmetry is estimated as
\begin{align}
\frac{n_{B-L}}{s} \simeq 
\dfrac{15}{4 \pi^2 g_{*s}} & \dfrac{ (\Delta_B-\Delta_L)(\mu_B(T_{\rm
 dec}) \Delta_B+\mu_L(T_{\rm dec}) \Delta_L) B^2L^2}{T_{\rm dec}(\Delta_B^2 L^2+\Delta_L^2 B^2)}\times \left\{
\begin{array}{ll}
1 & ~\text{for }~ T_{\rm dec}<T_R,\\
 \left(\dfrac{T_R}{T_{\rm dec}}\right)^5& ~\text{for }~ T_{\rm dec}>T_R.
\end{array} \right.  \label{bls}
\end{align}
After the decay of the flat direction, a part of the $B-L$ asymmetry is
converted to the baryon asymmetry through the sphaleron effect,
\begin{equation}
  \frac{n_B}{s}=\frac{8}{23} \frac{n_{B-L}}{s}.
\end{equation}
Thus, we have a successful baryogenesis scenario with appropriate
parameter choices such as reheating temperature or gravitino mass in the
context of the supersymmetric standard model \cite{arXiv:hep-ph/0304102}.

\subsection{Spontaneous Cogenesis \label{sec:2.2}}

We present a scenario of spontaneous cogenesis in the supersymmetric
flat directions in the similar way as discussed in the previous
subsection. For this purpose, we introduce an additional $U(1)$ symmetry
denoted as $U(1)_D$, which is responsible for the stability of dark
matter, and a pair of chiral supermultiplets $\Psi$ and ${\bar \Psi}$
whose $U(1)_D$ charge are $1$ and $-1$, respectively. These fields
construct a mass term,
\begin{equation}
W_{\rm mass}=M_\Psi \Psi {\bar \Psi}. 
\end{equation}
We also assume that the standard model fields are neutral under $U(1)_D$
and the superpotential of the system can be expressed as $W=W_{\rm
MSSM}+W_{\rm mass}$ in the renormalizable limit, where $W_{\rm MSSM}$ is
the superpotential for the MSSM.

\subsubsection{The case with only a $B$-$L$-$D$ mixing interaction \label{sec:genmix}}

Let us first consider a case that there is only a $B$-$L$-$D$ mixing
interaction. Denoting the amount of $B, L$ and $D$ violation as
$\Delta_{B}, \Delta_{L}$ and $\Delta_{D}$ for this interaction, produced
asymmetries must satisfy the following relations: 
\begin{equation}
\frac{n_B}{\Delta_B}=\frac{n_L}{\Delta_L}=\frac{n_D}{\Delta_D}, 
\end{equation}
which are equivalent to relations on the number densities of $m$-th field, 
\begin{equation}
\sum_m \Xi_m^{(1)} n_m = \sum_m \Xi_m^{(2)} n_m=0, 
\end{equation}
with
\begin{equation}
\Xi_m^{(1)} \equiv D_m \Delta_L-L_m \Delta_D, \quad 
\Xi_m^{(2)} \equiv D_m \Delta_B-B_m \Delta_D. \label{Xifor1}
\end{equation}
These relations are also rewritten in terms of chemical potential as
\begin{equation}
\Xi^{(1)} \cdot \mu = \Xi^{(2)} \cdot \mu =0. \label{constmix}
\end{equation}
Then, the physical chemical potential of $m$-th field can be expressed as
\begin{equation}
\mu_m = {\bar \mu}_m - \beta_1 \Xi_m^{(1)} - \beta_2 \Xi_m^{(2)},
\end{equation}
with the numerical parameters $\beta_1$ and $\beta_2$ given by
\begin{align}
\beta_1&=\frac{-\Delta_B^2 \mu_L L^2 D^2 -\Delta_D^2 \mu_L L^2 B^2 +\Delta_B \Delta_L \mu_B B^2 D^2}{\Delta_D (\Delta_B^2 L^2 D^2+\Delta_L^2 D^2 B^2+\Delta_D^2 B^2 L^2)},  \\
\beta_2&=\frac{-\Delta_L^2\mu_B B^2 D^2-\Delta_D^2 \mu_B B^2 L^2 +\Delta_B \Delta_L \mu_L L^2 D^2}{\Delta_D (\Delta_B^2 L^2 D^2+\Delta_L^2 D^2 B^2+\Delta_D^2 B^2 L^2)}. 
\end{align}
Then, the $B-L$ and $D$ asymmetries at the decoupling of the mixing
interaction are estimated as
\begin{align}
n_{B-L}(T_{\rm dec})&= \frac{T_{\rm dec}^2}{6} (B-L) \cdot \mu 
= \frac{ (\mu_B\Delta_B+\mu_L \Delta_L)(\Delta_B-\Delta_L) B^2 L^2
 D^2}{\Delta_B^2 L^2 D^2+\Delta_L^2 D^2 B^2+\Delta_D^2 B^2 L^2}
 \frac{T_{\rm dec}^2}{6}, \\
n_D(T_{\rm dec})&=\frac{T_{\rm dec}^2}{6}D \cdot \mu
= \frac{ (\mu_B\Delta_B+\mu_L \Delta_L) \Delta_D B^2L^2D^2}{\Delta_B^2
 L^2 D^2+\Delta_L^2 D^2 B^2+\Delta_D^2 B^2 L^2}  \frac{T_{\rm dec}^2}{6}, 
\end{align}
with $T_{\rm dec}$ being the decoupling temperature of the mixing
interaction. From these expressions, it is manifest that $\mu_B
\Delta_B+\mu_L\Delta_L \not=0$ is required for the generation of both
$B-L$ and $D$ asymmetries. In addition, $\Delta_B-\Delta_L \not = 0$ and
$\Delta_D\not=0$ are necessary for $B-L$ and $D$ asymmetries,
respectively.  The former condition implies that the mixing interaction
must violate $Q_+$ since the flat direction has $Q_+$ charge as a whole
and the derivative coupling acts as the chemical potential for
$Q_+$. The latter conditions reflect the fact that the mixing
interaction itself needs to violate $B-L$ and $D$ asymmetries.

Depending on the decoupling temperature, the present asymmetries are
estimated as
\begin{align}
\frac{n_{B-L}}{s}\simeq 
\dfrac{15}{4 \pi^2 g_{*s}} & \dfrac{ (\Delta_B-\Delta_L)(\mu_B(T_{\rm
 dec}) \Delta_B+\mu_L(T_{\rm dec}) \Delta_L)B^2L^2D^2}{T_{\rm dec}(\Delta_B^2 L^2 D^2+\Delta_L^2 D^2 B^2+\Delta_D^2 B^2 L^2)}\times \left\{
\begin{array}{ll}
1 & \text{for } T_{\rm dec}<T_R,\\
 \left(\dfrac{T_R}{T_{\rm dec}}\right)^5& \text{for } T_{\rm dec}>T_R,
\end{array} \right.  \label{nblsmix}\\
\frac{n_{D}}{s}\simeq 
\dfrac{15}{4 \pi^2 g_{*s}} & \dfrac{ \Delta_D(\mu_B(T_{\rm dec})
 \Delta_B+\mu_L(T_{\rm dec}) \Delta_L)B^2L^2D^2}{T_{\rm dec}(\Delta_B^2 L^2 D^2+\Delta_L^2 D^2 B^2+\Delta_D^2 B^2 L^2)}  \times \left\{
\begin{array}{ll}
1 & \text{for } T_{\rm dec}<T_R,\\
 \left(\dfrac{T_R}{T_{\rm dec}}\right)^5& \text{for } T_{\rm dec}>T_R.
 \end{array} \right.\label{ndsmix}
\end{align}
If annihilation process works sufficiently, only asymmetric part of $D$
charges, that is, the $\Psi$ (or $\bar{\Psi}$) particle remains,
which is responsible for the present dark matter. Since the ratio of the
$D$ asymmetry to the $B-L$ one is determined by the amounts of
violations of the mixing interaction,
\begin{equation}
\frac{n_{D}/s}{n_{B-L}/s}=\frac{\Delta_D}{\Delta_B-\Delta_L}, 
\end{equation} 
it is fixed after the freeze out of the mixing interaction. After the
decay of the flat direction, the sphaleron process converts a part of
the $B-L$ asymmetry into baryon asymmetry so that the present
baryon-to-entropy ratio is given by,
\begin{equation}
\frac{n_B}{s}=\frac{8}{23} \frac{n_{B-L}}{s}. \label{sphMSSM}
\end{equation}
Therefore, the DM mass given by 
\begin{equation}
M_\Psi \simeq 1.6 {\rm GeV} \times \frac{\Delta_B-\Delta_L}{\Delta_D}
\end{equation}
can explain not only the present dark matter abundance, 
\begin{equation}
\frac{\rho_{\rm DM}}{s}=\frac{M_\Psi n_D}{s} \simeq 4.1 \times 10^{-10} {\rm GeV}, \label{presDM}
\end{equation}
but also the present baryon-to-entropy ratio, 
\begin{equation}
\frac{n_B}{s}=(8.1-9.4) \times 10^{-11}. \label{presbary}
\end{equation}

\subsubsection{The case with both a $B$-$L$-$D$ mixing interaction and a
$B-L$ violating interaction \label{sec:genmixvio}}

Let us now consider a case with a $B-L$ violating interaction as well as
a $B$-$L$-$D$ violating (mixing) interaction, which can accommodate a
TeV scale dark matter model. Denoting the amounts of the charge
violation as $\Delta_{B1}$ and $\Delta_{L1}$ for the $B-L$ violating
interaction and $\Delta_{B2}, \Delta_{L2}$ and $\Delta_D$ for the
$B$-$L$-$D$ mixing interaction, respectively, the $B, L$ and $D$
asymmetries satisfy the following relation: 
\begin{equation}
\sum_m {\bar \Xi}_m n_m = \Delta_{L1}\Delta_D n_B -\Delta_{B1}\Delta_D n_L - (\Delta_{B2} \Delta_{L1}-\Delta_{B1}\Delta_{L2})n_D=0, \label{constboth}
\end{equation}
where ${\bar \Xi}_m$ is defined as 
\begin{equation}
{\bar \Xi}_m \equiv\Delta_{L1}\Delta_D B_m - \Delta_{B1}\Delta_D  L_m - (\Delta_{B2} \Delta_{L1}-\Delta_{B1}\Delta_{L2}) D_m, \label{Xibar}
\end{equation}
if there are no other symmetry breaking interactions. This relation is
rewritten in terms of chemical potential as
\begin{equation}
{\bar \Xi} \cdot \mu =0. 
\end{equation}
Then, the physical chemical potential of $m$-th field can be expressed as 
\begin{equation}
\mu_m={\bar \mu}_m-\beta {\bar \Xi}_m, 
\end{equation}
with the  numerical constant $\beta$ given by   
\begin{equation}
\beta=\frac{\left( \mu_B\Delta_{L1} B^2 -
	     \mu_L\Delta_{B1}  L^2 \right) \Delta_D}
{\Delta_{L1}^2\Delta_D^2  B^2+\Delta_{B1}^2\Delta_D^2 L^2+ (\Delta_{B2} \Delta_{L1}-\Delta_{B1}\Delta_{L2})^2 D^2}.  
\label{eq:beta}
\end{equation}
From this expression, the $B-L$ and $D$ asymmetries at the decoupling of
the mixing interaction\footnote{We assume that the $B-L$ violating
interaction decouples after the decoupling of the $B$-$L$-$D$ mixing
interaction since the symmetry the former breaks is smaller than that
the latter does.} are estimated as
\begin{align}
{n_B}(T_{\rm dec2})&= \frac{\left[(\mu_B \Delta_{B1}+\mu_L
 \Delta_{L1})\Delta_{B1} \Delta_D^2 L^2+ \mu_B (\Delta_{B2}
 \Delta_{L1}-\Delta_{B1}\Delta_{L2})^2
 D^2 \right]B^2}{\Delta_D^2(\Delta_{L1}^2 B^2+\Delta_{B1}^2 L^2)+
 (\Delta_{B2} \Delta_{L1}-\Delta_{B1}\Delta_{L2})^2 D^2}
 \frac{T_{\rm dec2}^2}{6}, \\
{n_L}(T_{\rm dec2})&= \frac{\left[(\mu_B \Delta_{B1}+\mu_L
 \Delta_{L1})\Delta_{L1} \Delta_D^2 B^2+ \mu_L (\Delta_{B2}
 \Delta_{L1}-\Delta_{B1}\Delta_{L2})^2 D^2 \right]L^2}
{\Delta_D^2 (\Delta_{L1}^2 B^2+\Delta_{B1}^2 L^2) + (\Delta_{B2}
 \Delta_{L1}-\Delta_{B1}\Delta_{L2})^2 D^2} \frac{T_{\rm dec2}^2}{6}, \\
n_{D}(T_{\rm dec2})&=\frac{\Delta_D(\mu_B\Delta_{L1}
 B^2-\mu_L\Delta_{B1} L^2)(\Delta_{B2}
 \Delta_{L1}-\Delta_{B1}\Delta_{L2})D^2}{\Delta_D^2 (\Delta_{L1}^2
 B^2+\Delta_{B1}^2 L^2)+ (\Delta_{B2}
 \Delta_{L1}-\Delta_{B1}\Delta_{L2})^2 D^2)}\frac{T_{\rm dec2}^2}{6}, 
\end{align}
with $T_{\rm dec2}$ being the decoupling temperature of the mixing
interaction.  Thus, it is manifest that the following three conditions
must be satisfied for successful dark matter genesis (cogenesis),
\begin{equation}
\Delta_D\not=0, \quad \Delta_{B2} \Delta_{L1}-\Delta_{B1}\Delta_{L2} \not=0, \quad \mu_B\Delta_{L1} B^2-\mu_L\Delta_{B1} L^2\not=0. 
\end{equation}
The first condition implies that $D$ violation is necessary for the
genesis of $D$ asymmetry. The second condition reflects the fact that,
only when $\Delta_{B2} \Delta_{L1}-\Delta_{B1}\Delta_{L2} \not =0$,
${\bar \Xi}_m$ depends on $D_m$ and the effective chemical potential of
$D$ charged particles arises. The last condition implies that the ratio
of $\mu_B B^2$ to $\mu_L L^2$ should not be equal to the ratio of
$\Delta_{B1}$ to $\Delta_{L1}$. The former represents the ratio of the
would-be baryon and lepton asymmetries generated by the effective
chemical potential $\bar{\mu}_m$ without the constraint
\eqref{constboth}. Note also that, if both of the ratios coincide, the
constant $\beta$ given in Eq.~(\ref{eq:beta}) vanishes. Thus, the last
condition requires that the $B$-$L$-$D$ mixing interaction as well as
the $B-L$ violating interaction must really work to generate B and L
asymmetries, which involves the generation of D asymmetry.

Depending on the decoupling temperature, the present $D$ asymmetry is
estimated in terms of number-to-entropy ratio as
\begin{align}
\frac{n_D}{s}\simeq 
\dfrac{15}{4 \pi^2 g_{*s}}  &\dfrac{ \Delta_D \left[ \mu_B(T_{\rm dec2})
 \Delta_{L1} B^2-\mu_L(T_{\rm dec2}) \Delta_{B1}L^2 \right](\Delta_{B2}
 \Delta_{L1}-\Delta_{B1}\Delta_{L2})D^2}{\left[ \Delta_D^2 (\Delta_{L1}^2
 B^2+\Delta_{B1}^2 L^2)+ (\Delta_{B2}
 \Delta_{L1}-\Delta_{B1}\Delta_{L2})^2 D^2 \right] T_{\rm dec2}} \notag \\
& \times \left\{
\begin{array}{ll}
1 & \text{for } T_{\rm dec2}<T_R,\\
 \left(\dfrac{T_R}{T_{\rm dec2}}\right)^5& \text{for } T_{\rm dec2}>T_R. 
 \end{array} \right. \label{twodark}
\end{align}
On the other hand, the $B-L$ violating interaction continues to generate
$B$ and $L$ asymmetries. Since the amounts of variation of these
asymmetries generated after the decoupling of the mixing interaction are
determined only by the violating interaction, $B$ and $L$ asymmetries
must satisfy the following relation,
\begin{align}
&\frac{n_B(T)-(a(T_{\rm dec2})/a(T))^3 n_B (T_{\rm dec2})}{n_L(T)-(a(T_{\rm dec2})/a(T))^3 n_L (T_{\rm dec2})}=\frac{\Delta_{B1}}{\Delta_{L1}} \notag \\
\Leftrightarrow& \Delta_{L1} n_B(T)-\Delta_{B1}n_L (T)=\left(\frac{a(T_{\rm dec2})}{a(T)}\right)^3\left(\Delta_{L1}n_B(T_{\rm dec2})-\Delta_{B1}n_L (T_{\rm dec2})\right) \notag \\
&=\left(\frac{a(T_{\rm dec2})}{a(T)}\right)^3\frac{(\mu_B \Delta_{L1}
 B^2-\mu_L \Delta_{B1} L^2)(\Delta_{B2}
 \Delta_{L1}-\Delta_{B1}\Delta_{L2})^2 D^2}{\left[ \Delta_D^2
 (\Delta_{L1}^2 B^2+\Delta_{B1}^2 L^2)+ (\Delta_{B2}
 \Delta_{L1}-\Delta_{B1}\Delta_{L2})^2 D^2 \right]} \frac{T_{\rm dec2}^2}{6} \notag \\
&\equiv C(T, T_{\rm dec2}). \label{ctt}
\end{align}
This relation is rewritten in terms of the chemical potential as 
\begin{equation}
{\tilde \Xi}\cdot \mu = \frac{6}{T^2}  C(T, T_{\rm dec2}), 
\end{equation}
with 
\begin{equation}
{\tilde \Xi}_m \equiv \Delta_{L1}B_m-\Delta_{B1} L_m. \label{Xitilde}
\end{equation}
Then, the chemical potential of $m$-th field can be expressed as
\begin{equation}
\mu_m={\bar \mu}_m-{\tilde \beta}{\tilde \Xi}_m,  
\end{equation}
with the numerical constant ${\tilde \beta}$ given by
\begin{equation}
{\tilde \beta}=\frac{(\mu_B \Delta_{L1} B^2-\mu_L \Delta_{B1} L^2)-6 C(T,T_{\rm dec2})/T^2}{\Delta_{L1}^2 B^2+\Delta_{B1}^2 L^2}. 
\end{equation}
The $B-L$ asymmetry at the decoupling of the $B-L$ violating
interaction, $T=T_{\rm dec1}$, is estimated as
\begin{equation}
n_{B-L}(T_{\rm dec1})=(\Delta_{B1}-\Delta_{L1})(\mu_B \Delta_{B1}+\mu_L
\Delta_{L1})\frac{B^2L^2}{B^2\Delta_{L1}^2+L^2\Delta_{B1}^2}\frac{T_{\rm
dec1}^2}{6}+\frac{ \Delta_{L1} B^2+\Delta_{B1}L^2 }{\Delta_{L1}^2
B^2+\Delta_{B1}^2 L^2}C(T_{\rm dec1},T_{\rm dec2}).
\end{equation}
In the case of $C(T_{\rm dec1}, T_{\rm dec2})=0$,\footnote{This is an
appropriate approximation since the asymmetry $\Delta_{L1}
n_B-\Delta_{B1}n_L$ at the decoupling of the mixing interaction is
considerably suppressed by a factor of $(a(T_{\rm dec2})/a(T))^3$. Even
if $C(T, T_{\rm dec2})$ has a non-negligible value, the present
asymmetry is easily calculated by the same procedure.} depending on the
decoupling temperature, the present $B-L$ number-to-entropy ratio is
evaluated as
\begin{align}
\frac{n_{B-L}}{s}=\frac{15}{4 \pi^2 g_{*2}}&\frac{(\Delta_{B1}-\Delta_{L1})(\mu_B(T_{\rm dec1}) \Delta_{B1}+\mu_L (T_{\rm dec1}) \Delta_{L1})B^2L^2}{(B^2\Delta_{L1}^2+L^2\Delta_{B1}^2)T_{\rm dec1}} \notag \\
&\times \left\{
\begin{array}{ll}
1 & \text{for } T_{\rm dec1}<T_R,\\
 \left(\dfrac{T_R}{T_{\rm dec1}}\right)^5& \text{for } T_{\rm dec1}>T_R. 
 \end{array} \right.
\end{align}
After the decay of the flat direction, the sphaleron process switches on
and reconfigures the baryon asymmetry so that the present baryon
asymmetry is estimated as
\begin{align}
\frac{n_{B}}{s}=\frac{30}{23 \pi^2 g_{*2}}&\frac{(\Delta_{B1}-\Delta_{L1})(\mu_B(T_{\rm dec1}) \Delta_{B1}+\mu_L (T_{\rm dec1}) \Delta_{L1})B^2L^2}{(B^2\Delta_{L1}^2+L^2\Delta_{B1}^2)T_{\rm dec1}} \notag \\
&\times \left\{
\begin{array}{ll}
1 & \text{for } T_{\rm dec1}<T_R,\\
 \left(\dfrac{T_R}{T_{\rm dec1}}\right)^5& \text{for } T_{\rm dec1}>T_R. 
 \end{array} \right. \label{nbsvio}
\end{align}
The ratio of the $D$ asymmetry to the $B-L$ asymmetry is suppressed by a
factor of $\CO((T_{\rm dec1}/T_{\rm dec2})^6)$ apart from a numerical
factor, if $T_R<T_{\rm dec1}, T_{\rm dec2}$ and the temperature
dependence of the chemical potential is neglected. Therefore, the mass
of $\Psi$ given by
\begin{equation}
M_\Psi\sim 1 {\rm GeV} \times \left(\frac{T_{\rm dec2}}{T_{\rm dec1}}\right)^6
\end{equation}
can explain the present dark matter abundance \eqref{presDM} and the
baryon-to-entropy ratio \eqref{presbary} simultaneously. Thus, this
scenario can accommodate dark matter with weak scale mass if $T_{\rm
dec2}/T_{\rm dec1} \sim \CO(10^{1/2})$.

One may wonder whether the baryon and dark matter isocurvature
perturbation would be too large to be consistent with the present
observations. The isocurvature perturbations roughly can be estimated by
\begin{equation}
\frac{\delta n_{B(D)}}{n_{B(D)}}\simeq \frac{\delta{\dot a}_+}{{\dot a}_+} \simeq \frac{\delta a_+}{a_+} \simeq \frac{1}{2\pi}\left(\frac{H_{\rm inf}}{M_*}\right)^{(n-3/n-2)}, 
\end{equation}
with $H_{\rm inf}$ being the Hubble parameter during inflation.
Therefore, for large $M_*$ compared to the Hubble parameter during
inflation, the baryon and dark matter isocurvature perturbations are
significantly suppressed.

\section{Application \label{sec:3}}

Here we consider concrete examples of spontaneous cogenesis in a flat
direction of the MSSM with $B-L=0$ and $B+L\not=0$. The AD mechanism
works only for a flat direction with $B-L$ charge, otherwise the
sphaleron process would wash out $B$ and $L$ asymmetries before the
electroweak phase transition.\footnote{If $Q$ balls are formed
and do not evaporate away before the EWPT, the AD mechanism is still
viable even for a flat direction with vanishing $B-L$ charge. Since $Q$
balls protect the asymmetries from the sphaleron process, both baryon
and lepton asymmetries can be generated in the Universe. However, in our
concrete models, $Q$ balls are not formed because the flat direction
quickly decays just after it starts the oscillation, as discussed
later.} On the other hand, spontaneous baryo/cogenesis can work even for
a flat direction without $B-L$ charge. Therefore, in order to avoid too
large asymmetry generated by the AD mechanism, we investigate the
cogenesis scenario along a flat direction without $B-L$ charge.

For definiteness, we choose the $QQQL$ flat direction parameterized as
\begin{equation}
Q_1^r=\frac{1}{\sqrt{2}}
\begin{pmatrix}
f e^{i\alpha_B/3}\\ 0
\end{pmatrix}, Q_2^b=\frac{1}{\sqrt{2}}
\begin{pmatrix}
0 \\ f e^{i\alpha_B/3}
\end{pmatrix}, Q_3^g=\frac{1}{\sqrt{2}}
\begin{pmatrix}
f e^{i\alpha_B/3}\\ 0
\end{pmatrix}, L_1 = \frac{1}{\sqrt{2}}
\begin{pmatrix}
0 \\ f e^{i\alpha_L}
\end{pmatrix}, 
\end{equation}
where $r,b$ and $g$ are color indices, and $f, \alpha_B$ and $\alpha_L$
are real parameters. In this example, all $f_i$'s defined in
Eq.~(\ref{eq:chii}) are identical and denoted by $f$, which yields
$\phi=f$ for the representative scalar field. Since the weak gauge
bosons become massive due to the large VEV of the flat direction, the
sphaleron process does not work until the decay of this flat
direction. For this flat direction, $\xi, Q^{\pm}_i$ and $v_a$ defined
in Eqs.~\eqref{qplus}, \eqref{qminus}, \eqref{xi} and \eqref{decayconst}
are given by
\begin{align}
\tan \xi & = 1, & Q^+_i&= \frac{B_i+L_i}{\sqrt{2}}, & Q^-_i&=\frac{-B_i+L_i}{\sqrt{2}}, & v_a&=\frac{\phi}{\sqrt{2}}. 
\end{align}
Since some of the fields become massive due to the direct coupling to
the flat direction when it acquires a large VEV, the quantities that
represent the light degrees of freedom of baryon, lepton, and dark
matter are estimated as
\begin{equation}
B^2 \equiv \sum_m \kappa_m g_m B_m^2=\frac{23}{2}, \quad L^2=\frac{81}{2}, \quad D^2=12, 
\end{equation}
respectively. The reason why half-integer baryon and lepton number
squared appear comes from the fact that some of massless eigenstates are
realized as mixed states of (s)quarks and (s)leptons.

In order to evaluate the effective chemical potential, we discuss the
evolution of the flat direction. Though the scalar potential along the
flat direction vanishes in the supersymmetric and renormalizable limit,
it is lifted up by a nonrenormalizable superpotential,
\begin{equation}
W_{\rm NR}=\frac{QQQL}{4M_*}, 
\end{equation}
with $M_*$ being a cutoff scale for this interaction and SUSY-breaking
effect. Assuming the negative Hubble induced mass squared, the flat
direction is destabilized at the origin and acquires a large VEV. After
inflation, the (negative) thermal logarithmic potential
Eq.~\eqref{negthlog} arises from thermal plasma that exists even in the
inflaton dominated era as a subdominant component of the Universe.  Note
that all the gauge fields become massive since the $QQQL$ flat direction
completely breaks the SM gauge groups. Thus, thermal logarithmic
potential coming from the two-loop finite temperature effects is
dominated by the effect of the running of the top Yukawa coupling with
the nonzero field value of the scalar fields.  Since the one-loop
correction to the top Yukawa coupling is negative, the thermal
logarithmic correction turns out to be negative \cite{kamadathesis}.
Then, the dynamics of the $QQQL$ flat direction follows the discussion
in Sec.~\ref{sec:2.1} with $n=4$. During the inflaton oscillation
dominated era, this potential has a minimum at
\begin{equation}
\phi_{\rm min} \sim \left\{
\begin{array}{ll}
(H M_*)^{1/2} & \text{for} \quad T>T_{\rm c}, \\
\alpha^{1/6} (T^2 M_*)^{1/3} & \text{for} \quad T<T_{\rm c}, 
\end{array} \right. \label{fminb}
\end{equation}
with the critical temperature $T_c$ given by
\begin{equation}
T_{\rm c} \equiv \alpha^{1/8} T_R^{3/4} M_G^{3/8} M_*^{-1/8}. 
\end{equation} 
After reheating, the negative Hubble induced mass squared
vanishes. Then, the potential minimum is determined by the balance of
the thermal logarithmic potential and the $F$ term, and is given by
\begin{equation}
\phi_{\rm min} \sim \alpha^{1/6} (T^2 M_*)^{1/3}.  \label{fmina}
\end{equation}
The flat direction follows this minimum until the mass term\footnote{Here we
assume the gravity mediated SUSY-breaking mechanism.} overwhelms the
thermal logarithmic potential at
\begin{equation}
T \sim \alpha^{-1/4} m_{3/2}^{3/4} M_*^{1/4}, \quad \phi \sim (m_{3/2} M_*)^{1/2}. 
\end{equation}
Below this temperature, the flat direction starts its oscillation around
the origin and decays quickly since its decay rate $\Gamma \sim m_{3/2}/
8 \pi$ is much larger than the Hubble parameter $H\sim T^2/M_G \sim
\alpha^{-1/2} m_{3/2} (f/M_G)$. Thus, $Q$ balls would not be formed
because of the quick decay of the flat direction.

While the NG boson $a_+$ slow-rolls, its velocity is estimated from
Eqs.~\eqref{adot1}, \eqref{fminb} and \eqref{fmina} as
\begin{align}
|{\dot a}_+|&=\frac{Q^+}{H v_a}M_A^4=\frac{a_{3/2} m_{3/2}}{8 M_* H} \phi_{\rm min}^3 \notag \\
&\simeq \left\{
\begin{array}{ll}
\dfrac{a_{3/2}}{8} m_{3/2} \phi_{\rm min} & \text{for} \quad T>T_{\rm c}, T_R,  \\
\dfrac{\alpha^{1/3} a_{3/2} m_{3/2} M_G T_R^2}{8 M_*^{1/3} T^{8/3}}\phi_{\rm min} & \text{for} \quad T_R<T<T_{\rm c}, \\
\dfrac{\alpha^{1/3}a_{3/2} m_{3/2} M_G}{8 M_*^{1/3} T^{2/3}} \phi_{\rm min}& \text{for} \quad T<T_R. 
\end{array} \right. \label{adot}
\end{align}
Then, the chemical potential $\bar{\mu}_m$ defined in
Eq.~(\ref{eq:effchem}) is given by
\begin{align}
{\bar \mu}_m &=-\frac{B_m+L_m}{\sqrt{2} v_a}{\dot a}_+ = B_m \mu_B+L_m \mu_L, \label{mubarQQQL}\\
|\mu_B|&=|\mu_L|\simeq \frac{a_{3/2}}{8}m_{3/2} \times \left\{
\begin{array}{ll}
1 & \text{for} \quad T>T_{\rm c}, T_R,  \\
\dfrac{\alpha^{1/3} M_G T_R^2}{ M_*^{1/3} T^{8/3}} & \text{for} \quad T_R<T<T_{\rm c}, \\
\dfrac{\alpha^{1/3} M_G}{ M_*^{1/3} T^{2/3}} & \text{for} \quad T<T_R, 
\end{array}\right. \label{mublQQQL}
\end{align}
where we have used Eqs.~\eqref{Leffm} and \eqref{adot}. The slow-roll
violating condition of the NG boson given in Eq.~\eqref{SR} is expressed
in terms of the temperature as
\begin{equation}
T<T_{\rm slow}\equiv \left\{
\begin{array}{ll}
(a_{3/2} m_{3/2} M_G T_R^2)^{1/4} & \text{for} \quad T_{\rm slow}>T_{\rm c},T_R, \\
\left(\dfrac{\alpha a_{3/2}^{3} m_{3/2}^{3} M_G^{6} T_R^{12}}{M_*}\right)^{1/20} & \text{for} \quad T_{\rm c}>T_{\rm slow}>T_R, \\
\left(\dfrac{\alpha a_{3/2}^{3} m_{3/2}^{3} M_G^{6} }{M_*}\right)^{1/8}
 & \text{for} \quad T_R>T_{\rm slow}. 
\end{array} \right.
\end{equation} 
Equation \eqref{mublQQQL} is valid for $T>T_{\rm slow}$. Otherwise, they are
severely suppressed because the field $a_+$ oscillates rapidly
\cite{Dolgov:1994zq}.

\subsection{The case with only a $B$-$L$-$D$ mixing interaction}

First we consider a spontaneous cogenesis scenario with only a
$B$-$L$-$D$ mixing interaction discussed in Sec.~\ref{sec:genmix}.  As
an illustrative example, we consider the following superpotential,
\begin{equation}
W_{{\rm mix}}= \frac{\bu \bd \bd \Psi^2}{\Lambda^2}, \label{supermix}
\end{equation}
which induces a $B$-$L$-$D$ mixing interaction,
\begin{align}
{\cal L}_{\rm vio} &= \frac{1}{\Lambda^2}  {\tilde \bu}_{R}{\tilde \bd}_{R}{\bd}_{R} \Psi^2  +{\rm h.c.}, 
\end{align}
where the tilde represents the fermion component of a chiral multiplet.
Here we impose $Z_2$ symmetry on the DM sector. Some flavor and color
combinations of $\bu\bd\bd$ (e.g., $\bu_{R1}^b \bd_{R2}^g \bd_{R3}^r$)
remain massless even though the $QQQL$ flat direction acquires a large
VEV. Then, this mixing interaction is in thermal equilibrium at high
temperature. More specifically, the rate of the interaction is given by
$\Gamma \sim T^5/(8 \pi \Lambda^4)$, which leads to its decoupling
temperature,
\begin{equation}
T_{\rm dec1}=\left\{
\begin{array}{ll}
8 \pi\dfrac{\Lambda^4}{M_G T_R^2}&\text{for} \quad T_{\rm dec1}>T_R, \\
(8 \pi)^{1/3}\dfrac{\Lambda^{4/3}}{M_G^{1/3}}&\text{for} \quad T_{\rm dec1}<T_R. 
\end{array}\right. 
\end{equation}
Since the amounts of $B, L$ and $D$ violation in this interaction are
given by
\begin{equation}
\Delta_{B}=-1, \quad \Delta_{L}=0, \quad \Delta_D=2,
\end{equation}
the operators $\Xi_m^{(1)}$ and $\Xi_m^{(2)}$ in Eq.~\eqref{Xifor1} are
written as
\begin{equation}
\Xi_m^{(1)}=-2L_m, \quad \Xi_m^{(2)} = -D_m - 2B_m.
\end{equation}
Then, the present $B-L$ and $D$ asymmetries are estimated from
Eqs. \eqref{nblsmix}, \eqref{ndsmix}, \eqref{mubarQQQL}, and
\eqref{mublQQQL} as
\begin{align}
\frac{n_{B-L}}{s}&=\frac{1035}{928 \pi^2 g_{*s}} \frac{a_{3/2} m_{3/2}}{T_{\rm dec}}\times \left\{
\begin{array}{ll}
\left(\dfrac{T_R}{T_{\rm dec}}\right)^5 & \text{for} \quad T_{\rm dec}>T_{\rm c}, T_R,  \\
\dfrac{\alpha^{1/3} M_G T_R^2}{ M_*^{1/3} T_{\rm dec}^{8/3}} \left(\dfrac{T_R}{T_{\rm dec}}\right)^5& \text{for} \quad T_R<T_{\rm dec}<T_{\rm c}, \\
\dfrac{\alpha^{1/3} M_G}{ M_*^{1/3} T_{\rm dec}^{2/3}} & \text{for} \quad T_{\rm dec2}<T_R, 
\end{array}\right. \\
\frac{|n_{D}|}{s}&=\frac{1035}{464 \pi^2 g_{*s}} \frac{a_{3/2} m_{3/2}}{T_{\rm dec}}\times \left\{
\begin{array}{ll}
\left(\dfrac{T_R}{T_{\rm dec}}\right)^5 & \text{for} \quad T_{\rm dec}>T_{\rm c}, T_R,  \\
\dfrac{\alpha^{1/3} M_G T_R^2}{ M_*^{1/3} T_{\rm dec}^{8/3}} \left(\dfrac{T_R}{T_{\rm dec}}\right)^5& \text{for} \quad T_R<T_{\rm dec}<T_{\rm c}, \\
\dfrac{\alpha^{1/3} M_G}{ M_*^{1/3} T_{\rm dec}^{2/3}} & \text{for} \quad T_{\rm dec}<T_R. 
\end{array}\right. 
\end{align} 
Here we have assumed that ${\dot a}_+<0$.

A part of the $B-L$ asymmetry is transferred to $B$ asymmetry through
the sphaleron process that works only after the $QQQL$ flat direction
decay. The present baryon asymmetry of the Universe is, then, estimated
as
\begin{equation}
\frac{n_B}{s}=\frac{8}{23}\frac{n_{B-L}}{s}\simeq \frac{45}{116 \pi^2 g_{*s}} \frac{m_{3/2}}{T_{\rm dec}}  \times \left\{
\begin{array}{ll}
\left(\dfrac{T_R}{T_{\rm dec}}\right)^5 & \text{for} \quad T_{\rm dec}>T_{\rm c}, T_R,  \\
\dfrac{\alpha^{1/3} M_G T_R^2}{ M_*^{1/3} T_{\rm dec}^{8/3}}\left(\dfrac{T_R}{T_{\rm dec}}\right)^5 & \text{for} \quad T_R<T_{\rm dec}<T_{\rm c}, \\
\dfrac{\alpha^{1/3} M_G}{ M_*^{1/3} T_{\rm dec}^{2/3}} & \text{for} \quad T_{\rm dec}<T_R. 
\end{array}\right. 
\end{equation}

The observed baryon asymmetry given in Eq.~\eqref{presbary} can be
explained with the parameters, for example, $T_R\simeq T_{\rm dec}
\simeq 10^{12}$ GeV, $m_{3/2} \simeq 4.5 \times 10^5$ GeV, and $M_*>2
\times 10^{27}$ GeV.  Notice that the conditions $T_R>T_{\rm c}, T_{\rm
slow}$ are satisfied for these parameters. If annihilation of the $\Psi$
fields occurs sufficiently (this issue is discussed later), the $\bar{\Psi}$
field with the mass of
\begin{equation}
M_\Psi \simeq 0.8 {\rm GeV}
\end{equation}
is also responsible for the present dark matter abundance given in
Eq.~\eqref{presDM}.

One may wonder such a high reheating temperature generates too many
lightest supersymmetric particles (LSPs) such as neutralinos that could
overclose the Universe or induces the gravitino problem. These problems
can be avoided if the mass of LSP is small enough, as is the case with
the axino LSP scenario. Such a small LSP mass is a natural assumption
since a light particle is required for the annihilation of DMs in the
first place.

\subsection{The case with both a $B$-$L$-$D$ mixing interaction and a $B-L$ violating interaction}

We consider a scenario with a $B-L$ violating operator given by
\begin{equation}
 {\cal L}_{\rm vio}=\frac{1}{\Lambda'^2} \bd_{R3}^3 {\tilde e}_{L2}
 {\tilde \nu}_{L3} + {\rm h.c.}, 
\end{equation}
as well as the $B$-$L$-$D$ mixing interaction given in
Eq.~\eqref{supermix}. The latter operator comes from a nonrenormalizable
superpotential,
\begin{equation}
W_{\rm vio}=\frac{\bd_3^3 L_2 L_3}{\Lambda'^2}, 
\end{equation}
with $\Lambda^\prime$ being a cutoff scale. As discussed in the previous
subsection, by taking adequate flavors and colors, all the fields
constituting of this interaction remain massless even though the $QQQL$
flat direction acquires a large VEV. The amounts of $B, L$ and $D$
violation of these interactions are given by
\begin{equation}
\Delta_{B1}=-1, \quad \Delta_{L1}=2, \quad \Delta_{B2}= -1, \quad
 \Delta_{L2}=0, \quad \Delta_D=2 ,
\end{equation}
which leads to the operators ${\bar \Xi}_m$ [Eq.~\eqref{Xibar}] and
${\tilde \Xi}_m$ [Eq.~\eqref{Xitilde}] respectively,
\begin{equation}
 {\bar \Xi}_m=4B_m+2L_m+2D_m, \quad {\tilde \Xi}_m=2 B_m+L_m.
\end{equation}

As discussed in Sec.~\ref{sec:genmixvio}, $D$ asymmetry is generated
thanks to the charge mixing/symmetry breaking interactions.  Depending
on the decoupling temperature, from Eqs.~\eqref{twodark},
\eqref{mubarQQQL} and \eqref{mublQQQL}, the present $D$ asymmetry is
estimated as
\begin{equation}
\frac{n_D}{s}=\frac{5715}{1576\pi^2 g_{*s}} \frac{a_{3/2} m_{3/2}}{T_{\rm dec2}}\times \left\{
\begin{array}{ll}
\left(\dfrac{T_R}{T_{\rm dec2}}\right)^5 & \text{for} \quad T_{\rm dec2}>T_{\rm c}, T_R,  \\
\dfrac{\alpha^{1/3} M_G T_R^2}{ M_*^{1/3} T_{\rm dec2}^{8/3}} \left(\dfrac{T_R}{T_{\rm dec2}}\right)^5& \text{for} \quad T_R<T_{\rm dec2}<T_{\rm c}, \\
\dfrac{\alpha^{1/3} M_G}{ M_*^{1/3} T_{\rm dec2}^{2/3}} & \text{for} \quad T_{\rm dec2}<T_R, 
\end{array}\right. 
\end{equation}
where we have assumed ${\dot a}_+>0$.

Assuming $T_{\rm dec1}<T_{\rm dec2}$, the factor $C(T,T_{\rm dec2})$
given in Eq.~\eqref{ctt} is evaluated as
\begin{equation}
C(T,T_{\rm dec2})=\left(\frac{a(T_{\rm dec2})}{a(T)}\right)^3 
 \frac{254 \mu_B(T_{\rm dec2}) T_{\rm dec2}^2}{197}, 
\end{equation}
with $\mu_B(T)=\mu_L(T)$. Then, the resultant $B-L$ asymmetry at the
decoupling temperature $T_{\rm dec1}$ is given by
\begin{equation}
n_{B-L}(T_{\rm dec1})= - \frac{1863 \mu_B(T_{\rm dec1})}{692}T_{\rm
dec1}^2-\frac{35}{173}C(T_{\rm dec1},T_{\rm dec2}).  \label{nblvio}
\end{equation}
Since $a^3 \mu(T) T^2$ increases rapidly as $T$ decreases, the second
term in Eq.~\eqref{nblvio} is negligible. Then, depending on the
decoupling temperature, from Eqs.~\eqref{nbsvio} and \eqref{mublQQQL},
the present $B$ asymmetry is estimated as
\begin{equation}
\frac{n_{B}}{s} \simeq  \frac{3645}{1384 \pi g_{*s}}   \frac{a_{3/2} m_{3/2}}{T_{\rm dec1}} \times \left\{
\begin{array}{ll}
\left(\dfrac{T_R}{T_{\rm dec1}}\right)^5 & \text{for} \quad T_{\rm dec1}>T_{\rm c}, T_R,  \\
\dfrac{\alpha^{1/3} M_G T_R^2}{ M_*^{1/3} T_{\rm dec1}^{8/3}}\left(\dfrac{T_R}{T_{\rm dec1}}\right)^5 & \text{for} \quad T_R<T_{\rm dec1}<T_{\rm c}, \\
\dfrac{\alpha^{1/3} M_G}{ M_*^{1/3} T_{\rm dec1}^{2/3}} & \text{for} \quad T_{\rm dec1}<T_R. 
\end{array}\right. 
\end{equation}
Here we have taken into account the sphaleron process that reconfigures
$B$ and $L$ asymmetries after the decay of the $QQQL$ flat
direction. The observed baryon asymmetry (Eq.~\eqref{presbary}) and the
dark matter abundance [Eq.~\eqref{presDM}] can be explained
simultaneously, for example, by the DM mass given by
\begin{equation}
m_\Psi \simeq 1.6  {\rm TeV}, 
\end{equation}
with the parameters, $T_R\simeq T_{\rm dec1}\simeq 10^{12}~{\rm GeV},
m_{3/2} \simeq 10^5~{\rm GeV} $, $M_*>5 \times 10^{27}$~GeV, and $T_{\rm
dec2} \simeq 3 \times 10^{12}~{\rm GeV}$, if the annihilation process
works sufficiently.

Note that the condition $T_{\rm dec1} \lesssim T_{\rm dec2}$ is a
natural assumption because they are both dimension 6 operators and hence
their cutoff scales have similar magnitudes with $\Lambda^\prime
\lesssim \Lambda$.  Operators whose dimension is less than 6 would
decouple at higher temperature in the inflaton oscillation dominated era
and at lower temperature in the radiation dominated era.  Thus, such
interactions may never be in thermal equilibrium in the cosmic history
and hence we do not consider them.

\subsection{Dark matter annihilation}

In order for a cogenesis scenario to be successful, the symmetric part
of DM must be annihilated enough at thermal freeze-out. Such an
annihilation process works significantly when its annihilation cross
section satisfies $\langle \sigma v_{\rm rel}\rangle>10^{-9} {\rm
GeV}^{-2}$ \cite{kolbturner,Iminniyaz:2011yp}.  Here we show that this
issue is accomplished by embedding our scenario in the
NMSSM, which can explain
the origin of the dark matter mass as well. Instead of giving the mass
term of dark matter $W_{\rm mass}=M_\Psi \Psi {\bar \Psi}$ and the $\mu$
term $W=\mu H_u H_d$ explicitly, we consider the following
superpotential,
\begin{equation}
\Delta W=\lambda_\Psi S \Psi {\bar \Psi} + \lambda_H SH_uH_d + \frac{\kappa }{3} S^3, \label{nmssm}
\end{equation}
where $S$ is an additional singlet field and $\lambda_\Psi, \lambda_H$
and $\kappa$ are numerical coefficients. Induced by the soft
SUSY-breaking effect, $S$ acquires a nonvanishing VEV, $\langle S
\rangle$, which generates the mass term of dark matter and the $\mu$
term,
\begin{equation}
M_\Psi = \lambda_\Psi \langle S \rangle, \quad \mu=\lambda_H \langle S
\rangle.
\end{equation}
For $\lambda_\Psi\simeq \lambda_H\simeq \CO(1)$ and $\langle S \rangle
\simeq 10^{2-3}$ GeV, we can obtain both the dark matter and the Higgs
masses with the electroweak scale. By decomposing the field $S$ into the
radial and the phase components $S=(s/\sqrt{2})e^{ia/s}$ with $s$
replaced by the VEV, $s=\sqrt{2} \langle S \rangle$, $\Psi$ and ${\bar
\Psi}$ can annihilate into the pair of $a$ field if its mass is smaller
than $M_{\Psi}$.\footnote{The field $a$ acquires the mass from the
$A$ term. Here we assume that it is smaller than the soft mass
$\simeq m_{3/2}$.}  This annihilation process occurs through the following
interaction,
\begin{equation}
\Delta {\cal L} = M_\Psi \Psi {\bar \Psi} e^{ia/s} \ni -\frac{M_\Psi}{s^2} \Psi {\bar \Psi}  a^2, 
\end{equation}
with the annihilation cross section given by
\begin{equation}
\langle \sigma v_{\rm rel} \rangle = \frac{1}{16 \pi} \frac{M_\Psi^2}{s^4}=\frac{\lambda_H^4 M_\Psi^2}{64 \pi \mu^4}. \label{ancr}
\end{equation}
For $M_\Psi \simeq s \simeq 1 {\rm TeV}$, we obtain
\begin{equation}
\langle \sigma v_{\rm rel} \rangle \sim 10^{-8} {\rm GeV}^{-2}, 
\end{equation}
which is large enough to annihilate the symmetric component of dark
matter. For $M_\Psi\simeq 1$ GeV, the annihilation cross section seems
too small to annihilate the symmetric part of dark matter
significantly. This problem can be avoided if we introduce a bare $\mu$
term,
\begin{equation}
\Delta W_H=\mu_0 H_u H_d, 
\end{equation}
in addition to the superpotential~\eqref{nmssm}. 
In this case, the observed $\mu$ term is expressed as 
\begin{equation}
\mu = \lambda_H \langle S \rangle + \mu_0. 
\end{equation}
Then, the annihilation cross section \eqref{ancr} is enhanced by a
factor of $(\mu/(\mu-\mu_0))^4$ and it can be large enough to annihilate
the symmetric part of DM sufficiently.

\section{Conclusion and Discussion\label{sec:4}}

In this paper, we have discussed an ADM model as an extension of the
spontaneous baryogenesis in the supersymmetric standard model. A
slow-rolling scalar field derivatively coupled to the matter current is
identified with a flat direction in the supersymmetric standard model,
which arises as a consequence of the Nambu-Goldstone's theorem. In this
setup, we have explicitly shown that the present abundances of the
baryon and the dark matter can be naturally explained by introducing a
nonrenormalizable $B$-$L$-$D$ mixing interaction and a few GeV dark
matter mass. Moreover, in the case that there is a $B-L$ violating
operator in addition to such a mixing interaction, the baryon asymmetry
can be slightly enhanced, which enables us to accommodate dark matter
with a TeV scale mass. On the other hand, such an enhancement of the
baryon asymmetry does not happen for a slow-rolling scalar field
derivatively coupled the dark matter current.

We have found that our model favors relatively high reheating
temperature and heavy gravitino mass, which may require some constraints
on the mass spectrum of SUSY particles or introducing another new
physical degree of freedom.  However, our model can be naturally
embedded in the NMSSM, which offers us a clue to solve this problem as
well as the annihilation mode for the symmetric part of dark matter. It
would be interesting to investigate its structure to satisfy the
requirements of our scenario in more detail.

%%%%%%%%%%%%%%%%%%%%%%%%%%%%%%%%%%%%
\section*{Acknowledgment}
%%%%%%%%%%%%%%%%%%%%%%%%%%%%%%%%%%%%

We would like to thank Wilfried Buchm\"uller, Kazuki Sakurai, Teruaki Suyama, and Fuminobu Takahashi for helpful
comments and discussions. This work is supported in part by JSPS
Grant-in-Aid for Scientific Research No.~21740187 (M.Y.).

%%%%%%%%%%%%%%%%%%%%%%%%%%%%%%%%%%%%

\end{document}